\pdfoutput=1
\documentclass{JINST}

\graphicspath{{./img/}}
\DeclareGraphicsExtensions{.pdf,.jpeg}

\title{A synchronous Gigabit Ethernet protocol stack for high-throughput UDP/IP applications}

\author{P. F\"odisch$^a$\thanks{Corresponding author.}, B. Lange$^a$, J. Sandmann$^a$, A. B\"uchner$^a$, W. Enghardt$^{b,c,d}$~ and P. Kaever$^a$\\
\llap{$^a$}Helmholtz-Zentrum Dresden - Rossendorf, Department of Research Technology,\\
Bautzner Landstr. 400, 01328 Dresden, Germany\\
\llap{$^b$}OncoRay - National Center for Radiation Research in Oncology, Faculty of Medicine and University Hospital Carl Gustav Carus, Technische Universit\"at Dresden,\\
Fetscherstr. 74, PF 41, 01307 Dresden, Germany\\
\llap{$^c$}Helmholtz-Zentrum Dresden - Rossendorf, Institute of Radiooncology\\
Bautzner Landstr. 400, 01328 Dresden, Germany\\
\llap{$^d$}German Cancer Consortium (DKTK) and German Cancer Research Center (DKFZ)\\
Im Neuenheimer Feld 280, 69120 Heidelberg, Germany\\

  E-mail: \email{p.foedisch@hzdr.de}}

\abstract{State of the art detector readout electronics require high-throughput data acquisition (DAQ) systems. In many applications, e.\,g. for medical imaging, the front-end electronics are set up as separate modules in a distributed DAQ. A standardized interface between the modules and a central data unit is essential. The requirements on such an interface are varied, but demand almost always a high throughput of data. Beyond this challenge, a Gigabit Ethernet interface is predestined for the broad requirements of Systems-on-a-Chip (SoC) up to large-scale DAQ systems. We have implemented an embedded protocol stack for a Field Programmable Gate Array (FPGA) capable of high-throughput data transmission and clock synchronization. A versatile stack architecture for the User Datagram Protocol (UDP) and Internet Control Message Protocol (ICMP) over Internet Protocol (IP) such as Address Resolution Protocol (ARP) as well as Precision Time Protocol (PTP) is presented. With a point-to-point connection to a host in a MicroTCA system we achieved the theoretical maximum data throughput limited by UDP both for 1000BASE-T and 1000BASE-KX links. Furthermore, we show that the random jitter of a synchronous clock over a 1000BASE-T link for a PTP application is below 60\,ps.}

\keywords{Gigabit Ethernet; Synchronous Ethernet; Field Programmable Gate Array (FPGA); High-throughput Data Acquisition (DAQ); User Datagram Protocol (UDP); Precision Time Protocol (PTP), 1000BASE-T, 1000BASE-KX, MicroTCA}
\begin{document}
\section{Introduction}
Distributed data acquisition systems are commonly spread over different fields of application in nuclear physics or medical imaging. Depending on the application, there are various requirements for the interconnections of the submodules. The main challenge for an interface is the user acceptance with respect to handling and interoperability of different device types. In addition, the data throughput of the interface is an important criterion for usability and should not limit the performance of the whole data acquisition (DAQ) system. Even though proprietary interfaces can fulfill these requirements, standardized technologies benefits from industry-proven components and are essential for reliable applications. A popular and well accepted specification is the IEEE 802.3 Standard for Ethernet \cite{bib1}. This standard specifies the physical layer used by the Ethernet. Until now, connections up to 100\,Gbit/s are specified and are going to be established by the industry. Nevertheless, for embedded systems a Gigabit Ethernet connection is the state of the art. A widespread technology is known as 1000BASE-T, which defines the 1\,Gbit/s Ethernet over twisted pair copper cables. The application of Gigabit Ethernet is not restricted to the use in Local Area Networks. It also finds its way into board-to-board applications. E.\,g. the backplane of a Micro Telecommunications Computing Architecture (MicroTCA) system should implement at least one port for an Ethernet connection \cite{bib2}, which is usually implemented as 1000BASE-KX on the physical layer. A link on the electrical backplane uses two differential pairs to establish a Gigabit Ethernet connection. With Gigabit Ethernet, the possibilities of applications range from a high speed data transfer to clock synchronization in a distributed DAQ system \cite{bib3}.
This work is related to the implementation and test of an embedded Gigabit Ethernet protocol stack for Field Programmable Gate Arrays (FPGA). With a versatile stack architecture, we will demonstrate the performance of high-throughput data transfers with the User Datagram Protocol (UDP) and clock synchronization over the Precision Time Protocol (PTP). Our aim is to investigate the maximum achievable data throughput with a FPGA-based System-on-Chip (SOC) as data source and a PC as receiver. For our application we need a high-throughput DAQ to cope with the gamma rate expected for prompt gamma imaging in ion beam therapy \cite{bib4}. This will be evaluated with a 1000BASE-T and 1000BASE-KX on a MicroTCA system. In addition, we will demonstrate the performance of a synchronized point-to-point connection as shown in \cite{bib5} with a Xilinx FPGA and different hardware for the physical layer.

\section{Requirements for an embedded Gigabit Ethernet protocol stack}
\subsection*{Physical layer}
\begin{figure}[ht]
\centering
\includegraphics[width=0.75\textwidth]{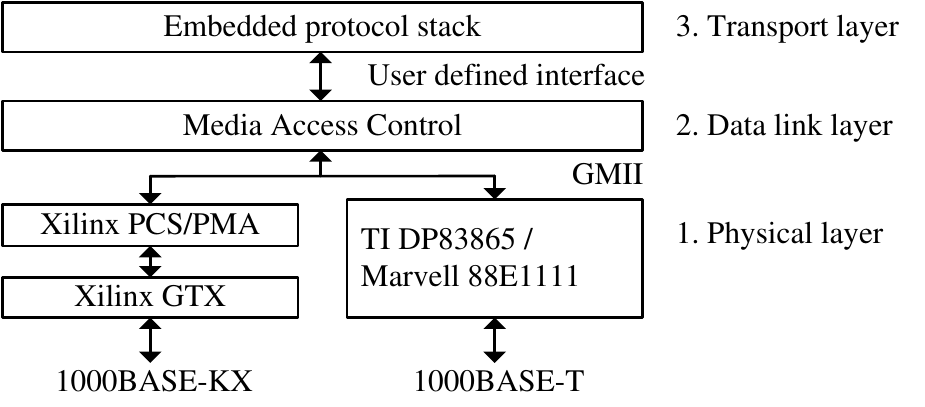}
\caption{Layer stack according to the OSI model for our embedded Gigabit Ethernet protocol stack. The higher level protocols are implemented in the transport layer. For evaluation of a 1000BASE-T link we use the external ICs DP83865 from Texas Instruments and 88E1111 from Marvell. All other components are implemented in a single FPGA.}
\label{fig_layer_stack}
\end{figure}

The embedded Gigabit Ethernet protocol stack connects to the physical layer through the data link layer regarding the Open Systems Interconnections (OSI) model as shown in fig.~\ref{fig_layer_stack}.
Higher level functions shall be implemented above the embedded protocol stack in an application layer. The IEEE 802.3 Standard for Ethernet defines different types of copper based connections between two transceivers over the physical layer. For 1000BASE-T, it is proposed to use four pairs of cables for the signal transmission. The 1000BASE-KX technology uses two pairs for the transmission of data. The specific signaling and coding in the physical layer will be done with industry-proven integrated circuits (ICs). For the 1000BASE-T signal coding we will use an IC from Marvell (88E1111, \cite{phy_marvell}) and Texas Instruments (DP83865, \cite{phy_ti}). To access the physical layer according to the \mbox{1000BASE-KX} technology, we will use a GTX Transceiver of the Xilinx Kintex~7 FPGA \cite{phy_gtx} in combination with the "Xilinx 1G/2.5G BASE-X PCS/PMA Core" \cite{phy_pcspma}. All physical layer transceivers (PHYs) have a common interface to the overlying data link layer and its Media Access Control (MAC). The MAC connects to the PHY via the Gigabit Media Independent Interface (GMII). The MAC should be implemented in the FPGA. Due to clear demands on high data throughput and hardware, we do not intend to provide a compatibility to other PHYs with Reduced or Serial Gigabit Media Independent Interface (RGMII or SGMII) or even lower speeds as specified for 10BASE-T or 100BASE-T.

\subsection*{Data link layer}
The data link layer with respect to fig.~\ref{fig_layer_stack} contains the MAC and a management interface. The MAC controls the access to the PHY and transmits the data in an Ethernet packet. It processes the input and output signals of the GMII with a frequency of 125\,MHz. An Ethernet packet encapsulates the Ethernet frame by adding the preamble and the start frame delimiter (SFD). The MAC composes (and also decomposes) the Ethernet packet with 8\,bits per clock cycle (8\,ns) from the Ethernet frame. This is essential for a maximum line rate of 1\,GBit/s. The standard for Ethernet demands that two consecutive Ethernet packets are separated by the interframe gap (IFG) for at least 96 bit times (96\,ns).\\
Usually all PHYs provide a management interface for the configuration of their internal register banks. The Media Dependent Input Output (MDIO) interface is used for a basic link configuration (e.g. autonegotiation advertisement).

\subsection*{Transport layer}
\label{sec_transport_layer}
The transport layer shall provide a stack for higher-level protocols encapsulated in the Ethernet frame. Its architecture must be easily extensible for any desired protocol in the layer stack. We target a maximized data throughput from the application layer for UDP. The theoretical data throughput of the UDP with a payload of 1472\,Byte, which corresponds to a Maximum Transfer Unit (MTU) of 1500\,Byte for the Ethernet frame, is 114.09\,MiB/s. If the host supports jumbo frames with a MTU of 9000\,Byte, the maximum data throughput is increased to 118.34\,MiB/s. The embedded protocol stack should not limit the frame size of an Ethernet packet. Although various implementations of Gigabit Ethernet protocol stacks (\cite{udp_ref1}, \cite{udp_ref2}, \cite{udp_ref3}, \cite{udp_ref4}, \cite{udp_ref5}, \cite{udp_ref6}, \cite{udp_ref7}, \cite{udp_ref8}, \cite{udp_ref9}, \cite{udp_ref10}, \cite{udp_ref11}, \cite{udp_ref12}, \cite{udp_ref13}) are published, there exists no solution which achieves the theoretical maximum data throughput with UDP. Only \cite{udp_ref4} reached maximum performance with a TCP/IP processor. A comparison of slice logic resources as it is done in \cite{udp_ref1}, \cite{udp_ref5}, \cite{udp_ref6}, \cite{udp_ref7}, \cite{udp_ref9} and \cite{udp_ref11} is not our intention, because each implementation is based on a different FPGA architecture. Whereas slice logic utilization is an important design criterion, it varies in accordance of generic configurations (e.g. FIFO depths) as well as the supported features (e.g. checksum calculations). Thus, a comparison to other implementations without the context of the application is difficult.
In order to provide all the necessary functionality, we need a protocol stack that serves Address Resolution Protocol (ARP), Internet Control Protocol (ICMP), Precision Time Protocol (PTP) and UDP with the focus on maximum data throughput. A protocol's header should be partially configurable by an user interface but also calculated automatically (e.\,g. length fields). All stack layers support a checksum calculation if it is required by the protocol. In terms of L\"ofgren's classification proposed in \cite{udp_ref1}, our requirements belong to a "Medium UDP/IP" core.

\section{Implementation}
\subsection{System overview}
\begin{figure}[ht]
\centering
\includegraphics[width=\textwidth]{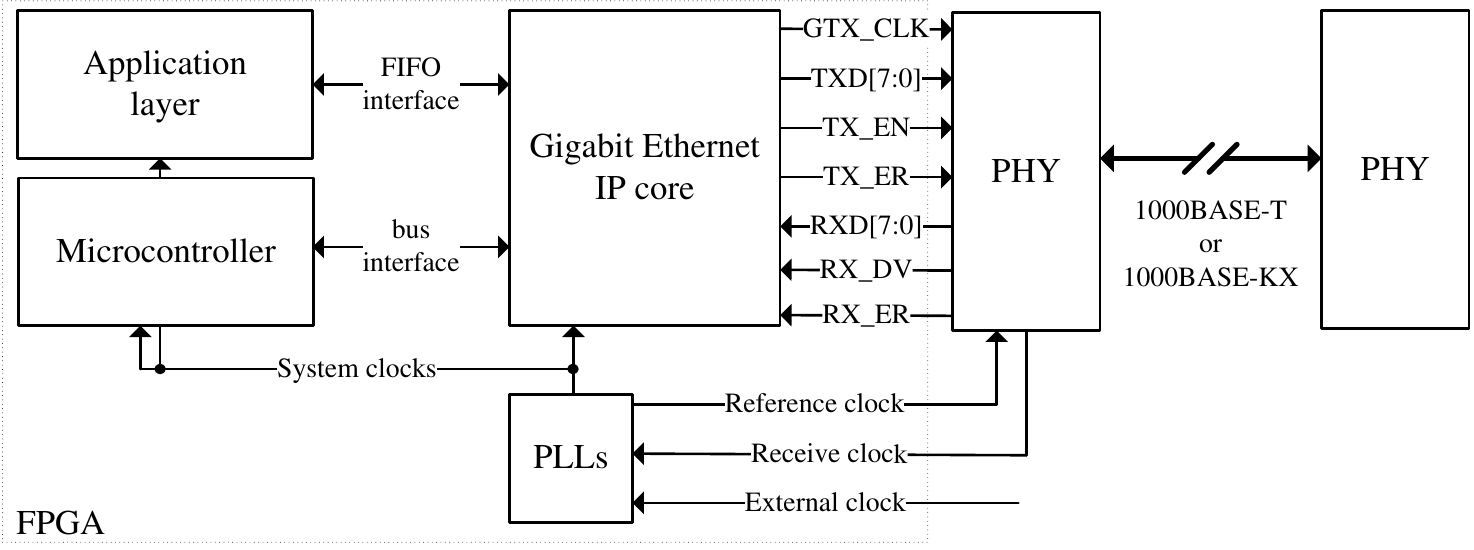}
\caption{An overview of the SoC with the embedded Gigabit Ethernet IP core and its interfaces. In case of the 1000BASE-KX implementation, the PHY will be included in the FPGA.}
\label{fig_system_overview}
\end{figure}

Our implementation is designed as Intellectual Property (IP) core with the hardware description language VHDL. It includes the MAC as well as an embedded protocol stack. For the \mbox{1000BASE-KX} implementation the PHY is already included in the FPGA. As shown in fig.~\ref{fig_system_overview}, the Gigabit Ethernet IP core gets its data from the application layer through a common First-In-First-Out (FIFO) interface.
The asynchronous FIFO is designed to operate at frequencies of 125\,MHz with a bit width of 32\,bit and stores at least the payload of one packet. It is used to stream the application data with high throughput into the core's transport layer (UDP). Another interface to the core is built with an 32\,bit microcontroller \cite{zylin_cpu}. The microcontroller with its bus-interface limits the data throughput for this interface far below a protocol's limit. So it is used for slow-control applications over UDP, ICMP and ARP. The MDIO management interface is also handled by the microcontroller and is not shown. Fig. \ref{fig_system_overview} shows the signals of the GMII and their directions between the MAC (embedded in the Gigabit Ethernet IP core) and the PHY. The same signals will be used for the 1000BASE-KX implementation with the embedded Xilinx PHY. The system clocks as well as the necessary clocks for the PHY will be generated by the FPGA's built-in phase-locked loop (PLL).

\subsection{Media Access Control}
\begin{figure}[ht]
\centering
\includegraphics[width=0.75\textwidth]{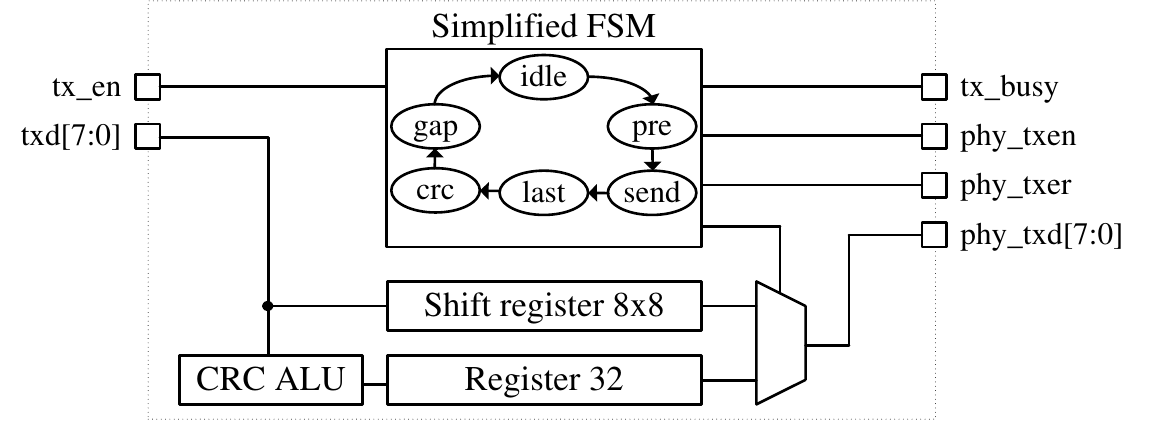}
\caption{Basic structure of the MAC for the transmission datapath. The output signals are connected to the GMII and the input is sourced by the transport layer.}
\label{fig_mac_tx}
\end{figure}

\begin{figure}[ht]
\centering
\includegraphics[width=\textwidth]{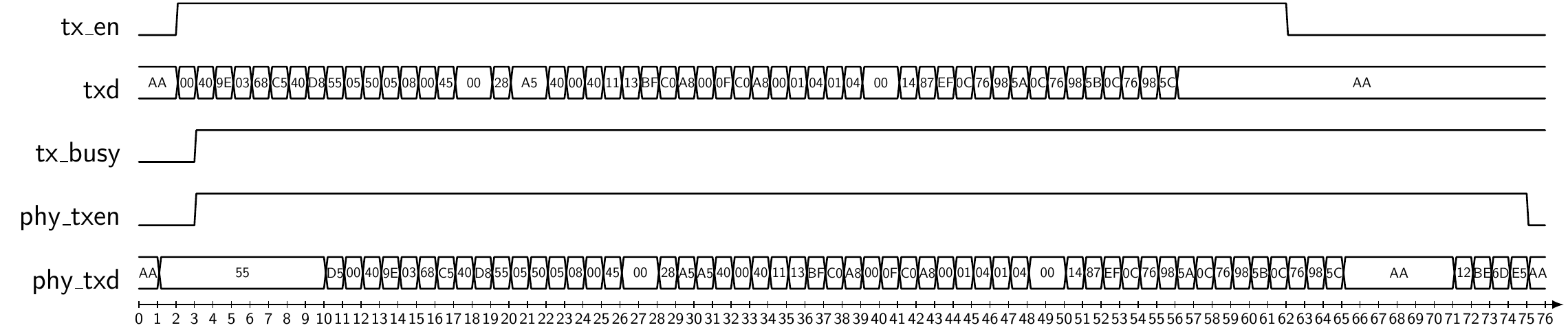}
\caption{An example of a composed Ethernet packet through the MAC layer for GMII. The transmission starts at position 3 with the preamble (signal "phy\_txd"). The MAC also adds the SFD (pos. 10), padding data (pos. 65-71) for a minimum payload length of 46\,Byte and the FCS (pos. 71-75). The IFG is controlled with the tx\_busy signal.}
\label{fig_mac_tx_chipscope}
\end{figure}

The functions of the MAC are restricted to the basic needs for interfacing the GMII. Fig. \ref{fig_mac_tx} shows the basic structure of the module for the transmission datapath of the MAC.
For the transmission datapath, it will compose the Ethernet packet with its preamble and SFD which are initially stored in a shift register. In the following states, data is passed through this register and the arithmetic logic unit (ALU) for the checksum calculation. Finally, the 32\,bit frame check sequence (FCS) is added at the end of the frame. The module's finite state machine (FSM) controls this dataflow and keeps the IFG at a programmable number of clock cycles. The module for the receiving datapath is built in the same way. It decomposes the Ethernet frame out of a received Ethernet packet and passes it to the transport layer. The MAC logic is capable of running at the speed of the transceiver clocks (125\,MHz). So there is no need for additional FIFOs for clock domain crossing. This results in a deterministic latency for the complete datapath from the transport layer to physical layer and vice versa. An example of a transmitted Ethernet packet is shown in fig.~\ref{fig_mac_tx_chipscope}. The waveforms are captured with an integrated logic analyzer (Xilinx Chipscope).

\subsection{Embedded protocol stack}
\begin{figure}[ht]
\centering
\includegraphics[width=\textwidth]{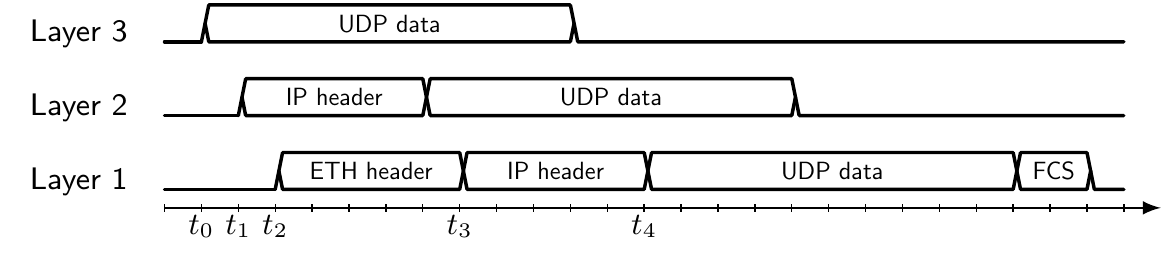}
\caption{An example of a dataflow through the stack layers driven by the "Data-Push" model}
\label{fig_data_push}
\end{figure}

\begin{figure}[ht]
\centering
\includegraphics[width=\textwidth]{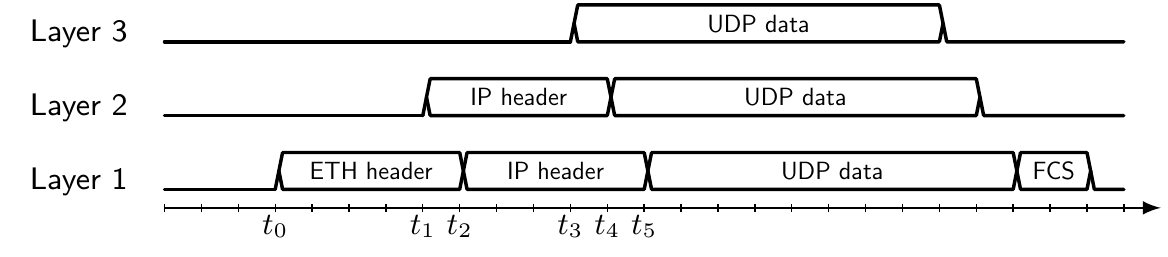}
\caption{An example of a dataflow through the stack layers driven by the "Data-Pull" model}
\label{fig_data_pull}
\end{figure}

\begin{figure}[ht]
\centering
\includegraphics[width=0.70\textwidth]{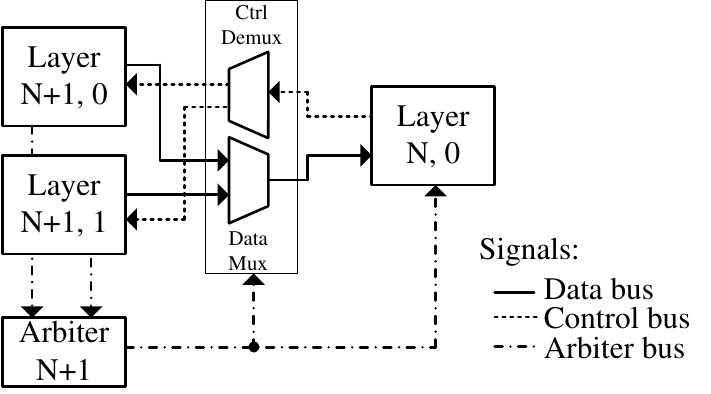}
\caption{Interconnections of layers with an arbiter and control logic. This architecture eliminates the need for redundant data buffers in a layer stack.}
\label{fig_module_interconnect}
\end{figure}

\begin{figure}[ht]
\centering
\includegraphics[width=\textwidth]{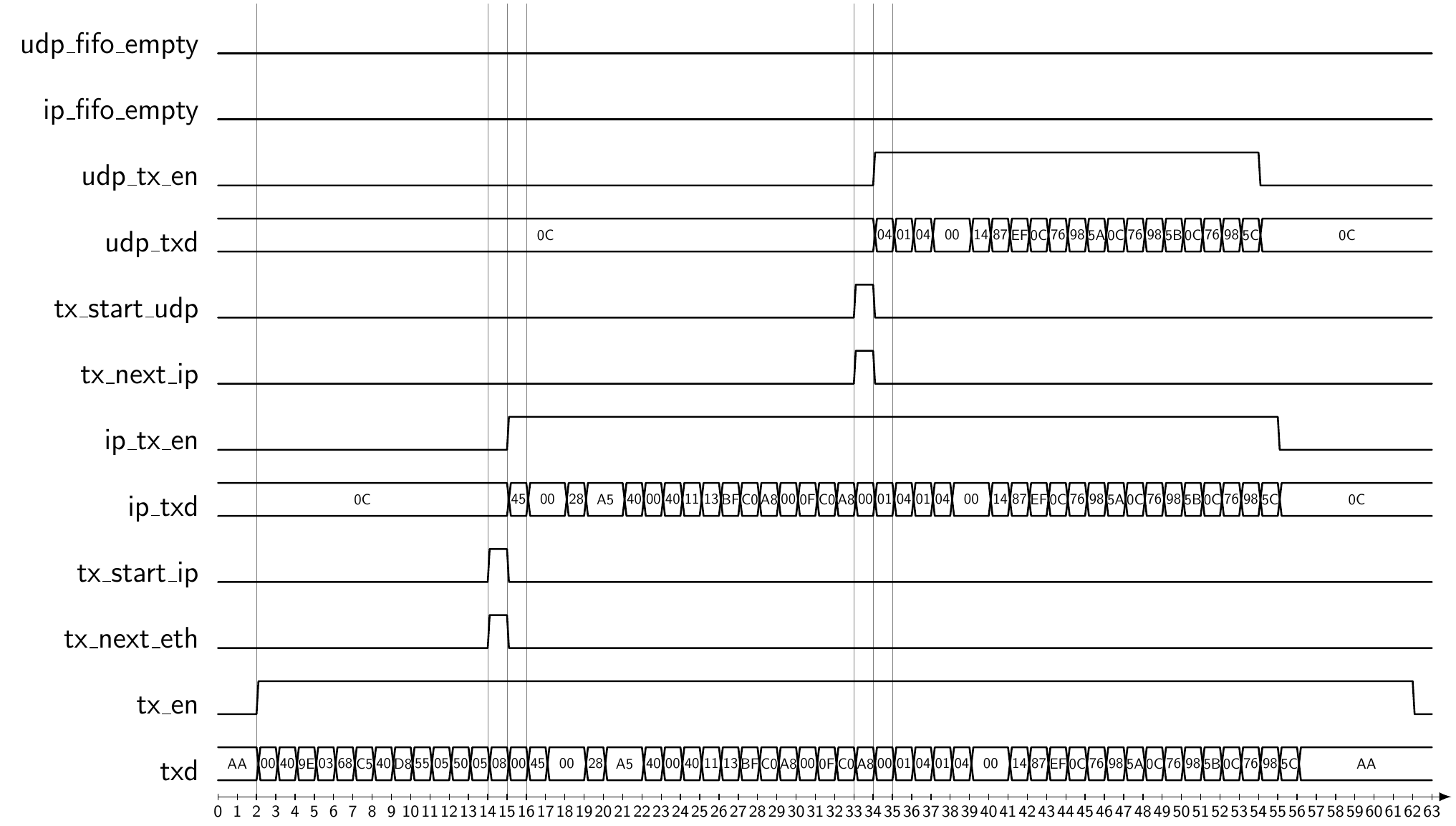}
\caption{Example of the dataflow for a UDP packet through the transport layers with the "Data-Pull" model.}
\label{fig_chipscope_tx_pull}
\end{figure}

With a look at the OSI reference model and its layers for a network communication, the stack architecture implies a dataflow from the top layer to the bottom layer. That means that the application passes its data from the transport layer to the data link layer until it is transmitted by the physical layer. So data will be "pushed" from the source to the sink and we call this dataflow as "Data-Push" model shown in fig.~\ref{fig_data_push}.
The scheme in fig.~\ref{fig_data_push} implies, that the application layer has valid data which is transported through the UDP layer (layer~3). The underlying IP layer (layer~2) will start its transmission with one clock cycle delay, beginning with its own data for the IP header. The data coming from the upper layer has to be buffered in the underlying layer, while this layer sends its own data. The same situation occurs when the IP layer passes its data to the Ethernet layer (layer~1). Finally, the dataflows initiated at time $t_0$ and $t_1$ are encapsulated at time $t_4$ and $t_3$ respectively. To keep this data valid for the latency during transmission time, data buffers are needed. As a consequence, a layer has to buffer at least the data of the overlying layer. One can also easily imagine the situation where two layers have valid data and pass it to a shared underlying layer. In this case, the number of data buffers doubles. A consistent data flow through all layers with the "Data-Push" model is handled with the appropriate number of data buffers. This model consumes additional memory for redundant data.\\
An alternative approach for a dataflow is shown in fig.~\ref{fig_data_pull}. We call this model as "Data-Pull" model.
In contrast to the "Data-Push" model from fig.~\ref{fig_data_push}, the dataflow is initiated by the low-level layer. The data of the overlying layers is just passed through a single  register stage at the time when it is encapsulated into the frame of the underlying layer. This reduces the amount of data buffers tremendously to a single register at each interconnection. In the example shown in fig.~\ref{fig_data_pull}, the latency from the UDP data in layer~3 to the time when it is encapsulated in layer~1 is reduced to two clock cycles (from $t_3$ to $t_5$). Each register stage in the underlying layer introduces one clock cycle delay. Of course the dataflow can be optimized to zero latency without additional register stages, but this will cause timing problems. Data buffers are needed in the application layers as well, but buffer redundancy in comparison to the "Data-Push" model is eliminated. The costs for this implementation are a simple arbiter and control logic and range far below those of the "Data-Push" approach. The basic scheme of the interconnections of layers is shown in fig.~\ref{fig_module_interconnect}.
All modules in the same layer N+1 pass their state to the arbiter logic. In the simplest case this is a FIFO state which indicates whether there is valid data to send or not. In case of valid data the arbiter decides which module of a layer is served first and passes this information to the underlying layer N. The module from layer N controls the dataflow of the overlying layer with its controlbus. After all, the data from layer N+1 is multiplexed to the receiving module in layer N. A real data transfer of the implemented "Data-Pull" model is shown in fig.~\ref{fig_chipscope_tx_pull}.
The example in fig.~\ref{fig_chipscope_tx_pull} shows at its initial clock cycle at position 1 that the UDP layer has valid data to send (signal "udp\_fifo\_empty" is low). In conjunction with the arbiter bus, the IP layer also reports that there is valid data to send (signal "ip\_fifo\_empty" is low). With this condition the Ethernet layer starts the transmission of data (signal "tx\_en" is high) at pos. 2. At position 14 the Ethernet layer pulls the data from the overlying layer by setting the signal "tx\_next\_eth" to high. The Ctrl Demux from the interconnection logic of the layers shown in fig.~\ref{fig_module_interconnect} switches this signal to the IP layer (signal "tx\_start\_ip" is high). At the next clock cycle (pos. 15), the IP layer transmits its data occurring with an additional delay of one clock cycle in the frame of the Ethernet layer (signal "txd", pos. 16). The IP layer encapsulates the application data from the UDP layer in the same way into its frame. This can be seen by the control signals "tx\_next\_ip" and "tx\_start\_udp" at position 33 and the UDP data (signal "udp\_txd") and the IP data (signal "ip\_txd") at pos. 34 and 35 respectively. Finally, the MAC composes the entire packet as shown in fig.~\ref{fig_mac_tx_chipscope}.

\subsection{Clock synchronization}
\begin{figure}[ht]
\centering
\includegraphics[width=\textwidth]{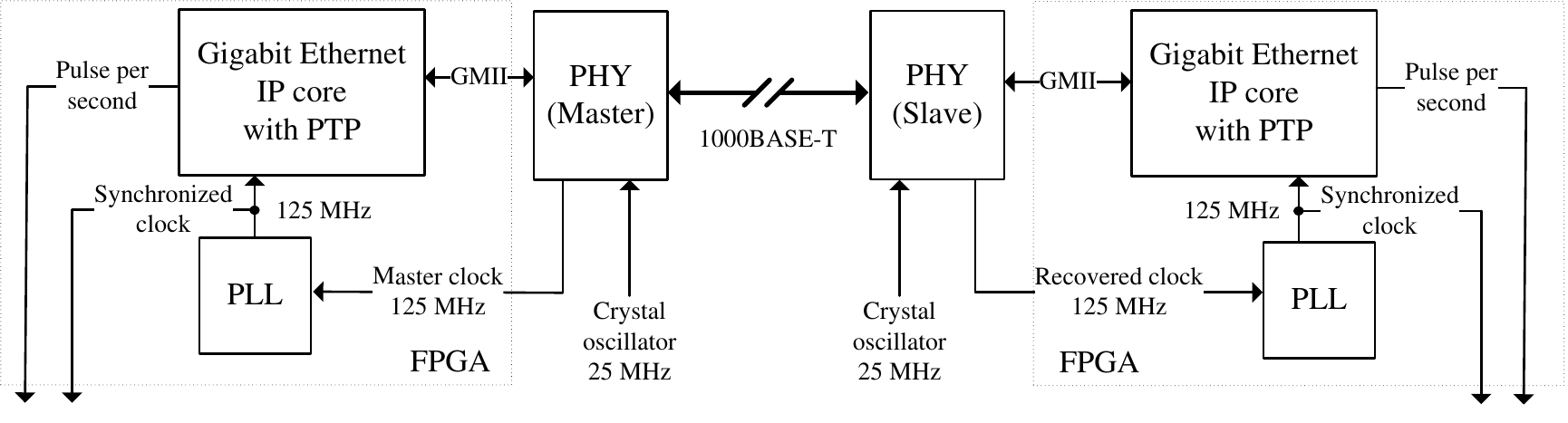}
\caption{Scheme of the clock synchronization through a point-to-point connection over a 1000BASE-T link. One PHY acts as master and embeds the clock reference into the datastream. The slave recovers a synchronized clock signal with a frequency of 125\,MHz. A PLL inside the FPGA is used to build up the clock tree. Our Ethernet IP core provides synchronized timestamps and a pulse per second for the test setup.}
\label{fig_clocking_scheme}
\end{figure}

An important issue in a distributed DAQ is a uniform clock distribution. Although a dedicated clock line is a simple and precise solution, it cannot be used for an absolute synchronization of all timestamps in the system. For this purpose an additional data signal for the transmission of a known timestamp reference is needed. The PTP offers the possibility to synchronize the timestamps over a data link. Additionally, a Gigabit Ethernet link has the property, that a transmission clock is embedded in the datastream, because the transferred data is synchronous to this reference clock. As a consequence, a receiver can recover this clock frequency. In a 1000BASE-T application, the slave recovers the master's clock out of the data stream. This task is done by the PHY (see fig.~\ref{fig_clocking_scheme}). So it is possible to synchronize the clock signals as well as the timestamps over a single Gigabit Ethernet link. It is also known that the clock offset from a master and a slave can not be corrected with PTP below a resolution of the 8\,ns (this corresponds to the transceiver clock of 125\,MHz) without a phase alignment of the clocks. An accurate implementation is already done with the White Rabbit project \cite{bib3}, but does not support a 1000BASE-T link by default. A synchronization over a 1000BASE-T link was done by \cite{bib5}. They achieved a precision of 180\,ps with the DP83865 from Texas Instruments and a FPGA from Altera. The limiting factor was the jitter of the FPGA's built-in PLL. Our implementation is based on Xilinx FPGAs with an improved jitter. So we want to determine the absolute precision which is achievable with these devices and different ICs for the physical layer. The implemented clocking scheme is shown in fig.~\ref{fig_clocking_scheme}. Each PHY is configured with the MDIO interface to act as a master or as a slave. During the autonegotiation procedure, these configurations are advertised.

\section{Measurements and results}
\begin{figure}[ht]
\centering
\includegraphics[width=0.75\textwidth]{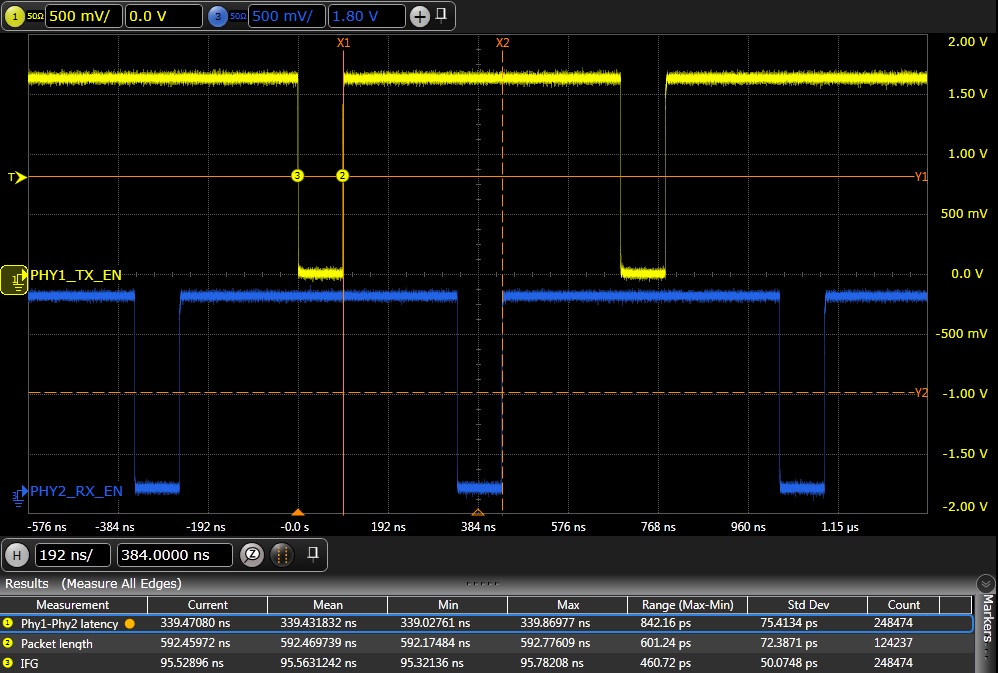}
\caption{The transmission enable signal (signal "PHY1\_TX\_EN") of the transmitting MAC and the receive enable signal (signal "PHY2\_RX\_EN") of the receiving PHY. The oscilloscope measurement verifies that the MAC keeps the IFG at 96\,ns while sending data with maximum throughput and constant latency. The packet length of 592\,ns corresponds to a UDP payload of 20\,Byte.}
\label{fig_ifg}
\end{figure}

For our performance test on a 1000BASE-T link we use the Xilinx evaluation board SP605 equipped with a Spartan~6 (LX45T) FPGA and the PHY 88E1111 from Marvell. We also use a FPGA Mezzanine Card (FMC) equipped with two PHYs from Texas Instruments (DP83865) attached to the SP605. The host is a MicroTCA crate equipped with an Advanced Mezzanine Card (AMC) CPU module from Concurrent Technologies (AM 900/412-42) and a MicroTCA Carrier Hub (MCH) from N.A.T. (NAT-MCH-PHYS). The CPU module provides two 1000BASE-T ports at the front and two 1000BASE-KX ports at the backplane. A 1000BASE-KX link to the CPU is established with a Kintex~7 (325T) on the HGF-AMC from DESY/KIT through the MicroTCA backplane and the switch from the MCH. The operating system on the host is Ubuntu.\\
To evaluate the MAC Layer and the latency of the entire stack, we measured its output signals on the GMII. A maximum throughput is achieved if the transmit enable signal (see "phy\_txen" in fig.~\ref{fig_mac_tx_chipscope}) is high all the time except the time for the IFG. A constant latency is achieved, if a transmission cycle and the arrival time of a packet at the receiver have a time deviation much smaller than a clock cycle. Both conditions could be experimentally verified, which indicates that the MAC layer is capable of transferring the maximum throughput with a constant latency (see fig.~\ref{fig_ifg}).
The measurement shown in fig.~\ref{fig_ifg} was captured with an oscilloscope during a transmission of UDP packets with a fixed payload of 20\,Byte. This test was chosen to verify a maximum throughput, a constant latency of the core (see measurement "packet length" and "IFG" in fig.~\ref{fig_ifg}) and the overall system latency between two PHYs (see measurement "Phy1-Phy2 latency" in fig.~\ref{fig_ifg}). For this setup we used the Marvell 88E1111 on both sides connected by a cable of 50\,cm length.

\subsection{Throughput performance}
\begin{figure}[ht]
\centering
\includegraphics[width=\textwidth]{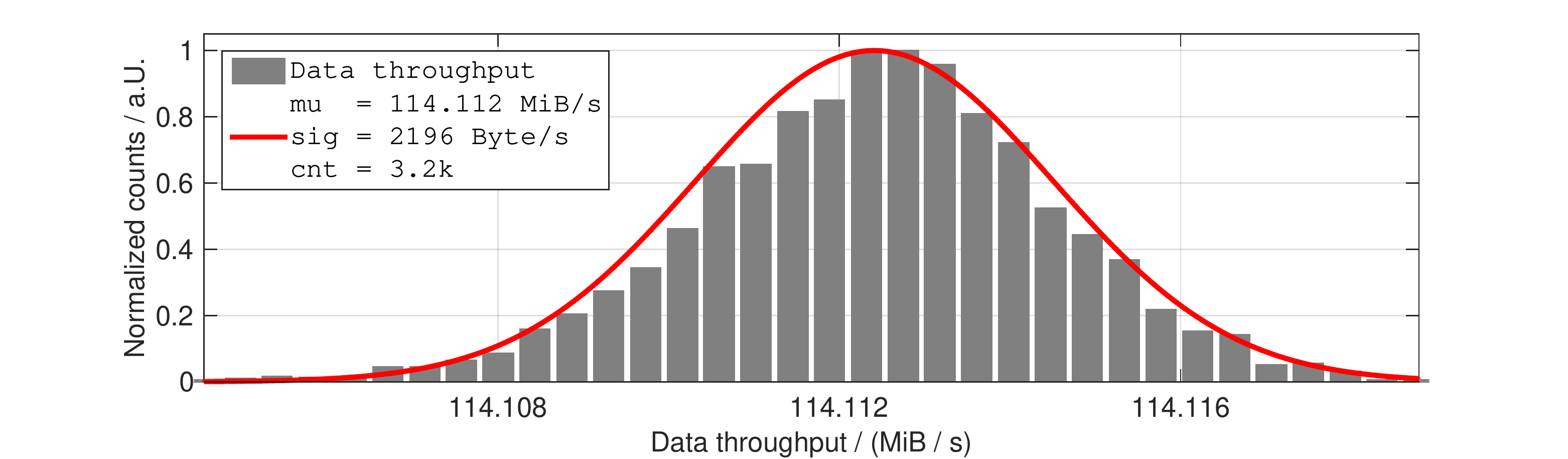}
\caption{Distribution of data throughput for 1472\,Byte UDP payload measured over 9\,hours.}
\label{fig_distri_throughput}
\end{figure}

To check the performance of our FPGA implementation with a 1000BASE-T PHY we directly established a point-to-point connection between the FPGA and the CPU module's front connector. The host serves a UDP socket where the incoming data throughput is measured. The data from the FPGA contains an increasing 32\,bit counter value which is used to identify a missing packet or a corrupted datastream. For this measurement the throughput of the UDP on a Gigabit Ethernet link is our reference. As mentioned in sec. \ref{sec_transport_layer}, this value is 114.09\,MiB/s for a payload of 1472\,Byte. If the payload is decreased, the data throughput decreases as well because of the increasing rate of protocol overhead. The table \ref{tab_throughput_ref} shows the achievable data throughput in dependence of the UDP payload. Additional overhead in the Ethernet packet limits the line rate.
\begin{table}[ht]
\caption{Theoretical data throughput in dependence of the payload of a UDP packet.}
\label{tab_throughput_ref}
\centering
\begin{tabular}{|l|l|l|}
\hline
\bfseries UDP payload / Byte & \bfseries Data throughput / (MiB/s) & \bfseries Line rate / (1 GBit/s) \\
\hline
8972 & 118.339 & 99.3 \%\\
\hline
1472 & 114.094 & 95.7 \%\\
\hline
1024 & 111.991 & 93.9 \%\\
\hline
512  & 105.597 & 88.6 \%\\
\hline
256  & 94.775  & 79.5 \%\\
\hline
\end{tabular}
\end{table}
Thus the Ethernet Standard limits the MTU per frame to 1500\,Byte (this results in a UDP payload of 1472\,Byte), it is also common to use jumbo frames with a MTU of 9000\,Byte (8972\,Byte UDP payload). Our implementation supports jumbo frames and this performance will also be evaluated with the MicroTCA host. The measurement of data throughput at the host requires also a measurement of time for the corresponding amount of bytes. Because Linux is no real-time operating system, this time measurements are above the nanosecond scale. But for an average estimation of data throughput this is sufficient. Our application measures the incoming bytes on the socket, checks if data is valid and prints out the error rate and the data throughput every ten seconds. The results of the data throughput tests with three different devices on the physical layer are shown in tab.~\ref{tab_result_throughput}. An example of a measurement of data throughput over 9\,hours is shown in fig.~\ref{fig_distri_throughput}. During this measurement all data were transferred without errors. The standard deviation of data throughput was 2196\,Byte/s. This is caused by uncertainties in the time measurement and the latency of packet buffering in the operating system.
\begin{table}[ht]
\caption{Measured data throughput in dependence of the UDP payload. The tests were performed with different hardware platforms. Data throughput is the mean value for more then 100\,s.}
\label{tab_result_throughput}
\centering
\begin{tabular}{|l|l|l|l|}
\hline
\bfseries Physical layer & \bfseries Hardware &  \bfseries UDP payload / Byte & \bfseries Data throughput / (MiB/s)\\
\hline
{1000BASE-T} & {TI DP83865}         & 8972 & 118.344\\
{1000BASE-T} & {Marvell 88E1111}    & 8972 & 118.345\\
{1000BASE-KX} & {Xilinx GTX}        & 8972 & 118.316\\
\hline
{1000BASE-T} & {TI DP83865}         & 1472 & 114.112\\
{1000BASE-T} & {Marvell 88E1111}    & 1472 & 114.097\\
{1000BASE-KX} & {Xilinx GTX}        & 1472 & 114.079\\
\hline
{1000BASE-T} & {TI DP83865}         & 1024 & 112.015\\
{1000BASE-T} & {Marvell 88E1111}    & 1024 & 111.995\\
{1000BASE-KX} & {Xilinx GTX}        & 1024 & 111.977\\
\hline
{1000BASE-T} & {TI DP83865}         & 512  & 105.643\\
{1000BASE-T} & {Marvell 88E1111}    & 512  & 105.641\\
{1000BASE-KX} & {Xilinx GTX}        & 512  & 105.619\\
\hline
{1000BASE-T} & {TI DP83865}         & 256  & 94.827\\
{1000BASE-T} & {Marvell 88E1111}    & 256  & 94.836\\
{1000BASE-KX} & {Xilinx GTX}        & 256  & 94.823\\
\hline
\end{tabular}
\end{table}
The results show an excellence performance up to the theoretical limit of the UDP data throughput. The values above this limits are caused by frequency uncertainties for the transmission clock. The reference values from tab.~\ref{tab_throughput_ref} are calculated at a clock frequency of 125\,MHz. A fixed deviation of that frequency and the mentioned lack of a precise time measurement on a Linux system can cause a data throughput value above the reference value. This tests also show the importance of an efficient host as receiver. If the host is not configured appropriately, packet losses will happen. In our configuration packet losses at the host receiver just occur at payloads smaller than approximately 256\,Byte. This is caused by the excessive load of more than 388\,kPackets/s.
As a second test we measured the performance of the ICMP protocol layer with ordinary Ping requests from the host to the FPGA device. A Ping request is processed by an interrupt routine with the microcontroller inside the FPGA. The host generated at least 1000 Ping requests at an interval of 200\,ms. The results are shown in tab.~\ref{tab_ping_result}.
\begin{table}[ht]
\caption{The results of the ICMP layer test with Ping requests. The host generated 1000 Ping requests and the RTT is measured.}
\label{tab_ping_result}
\centering
\begin{tabular}{|l|l|l|l|l|l|l|}
\hline
\bfseries Physical layer  & \bfseries Hardware & \bfseries Min / ms & \bfseries RTT mean / ms & \bfseries Max / ms & \bfseries Std.-dev / ms\\
\hline
{1000BASE-T}  & {TI DP83865}      & 0.203 & 0.315 & 0.444 & 0.067\\
{1000BASE-T}  & {Marvell 88E1111} & 0.189 & 0.307 & 0.547 & 0.066\\
{1000BASE-KX} & {Xilinx GTX}      & 0.740 & 1.091 & 1.441 & 0.162\\
\hline
\end{tabular}
\end{table}
The Ping requests were sent while the FPGA transmits UDP packets with maximum data throughput. The arbiter of the IP layer prioritizes the ICMP protocol, so that a parallel UDP data stream did not block the ICMP layer. Because of the MCH's switch for the 1000BASE-KX backplane links, the round-trip time (RTT) for a Ping request is higher than for a point-to-point connection. If the UDP layer application was turned off, the RTT of the backplane link was decreased to 0.252\,ms with 0.051\,ms standard deviation. The same test was done for the ARP layer, which has a higher priority than the IP layer in the protocol stack. The Linux host generated additional ARP request with the arping command, while the FPGA handles the ICMP and UDP layer. As a result, all request were served without losses with a RTT below 1\,ms for all hardware platforms.

\subsection{Synchronization}
\begin{figure}[ht]
\centering
\includegraphics[width=\textwidth]{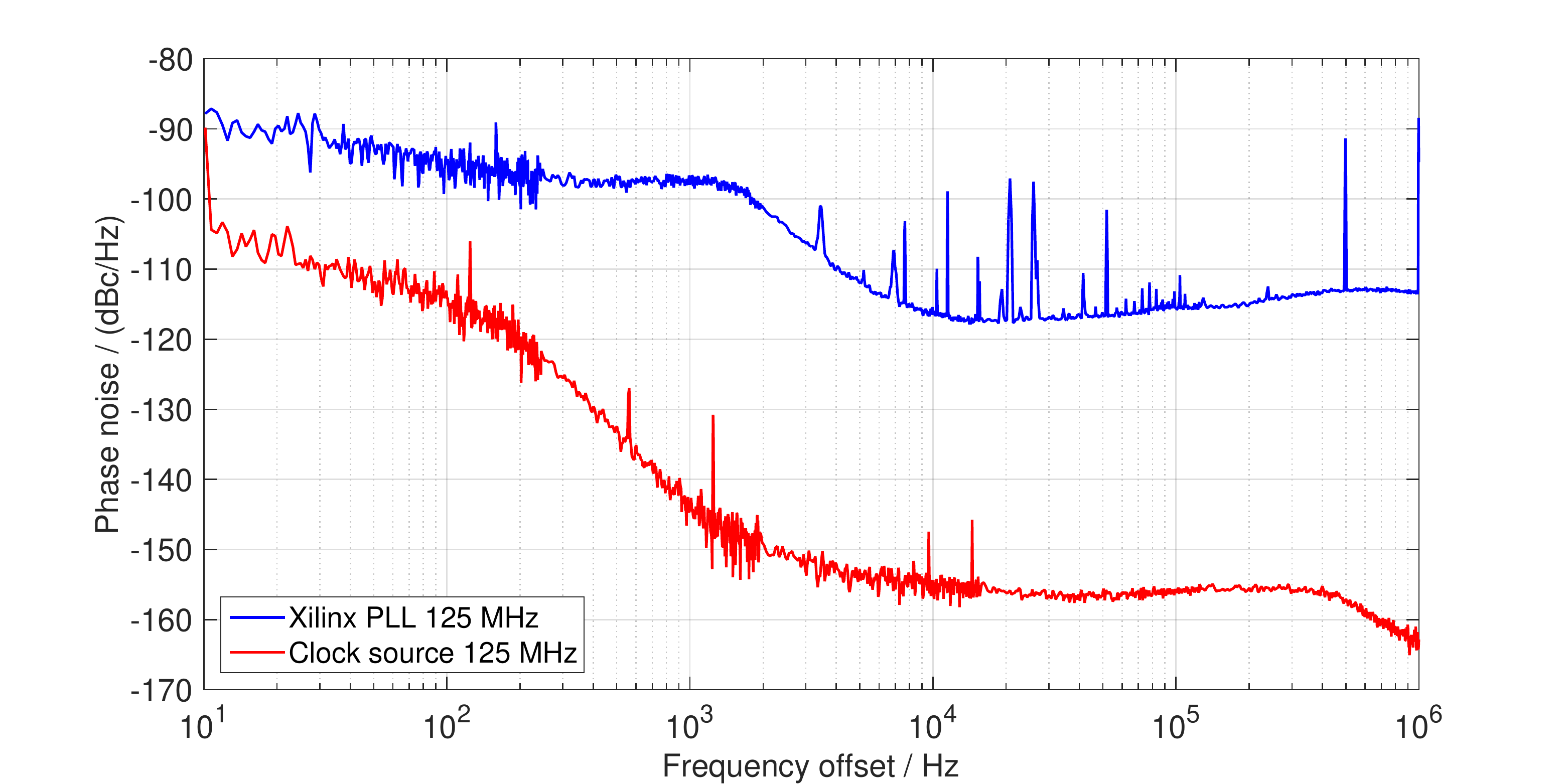}
\caption{A measurement of the phase noise of a 125\,MHz low jitter clock source (red) which sources a PLL in a Spartan~6 LX45T. The corresponding output of the PLL (blue) has an integrated phase noise of 6.47\,ps in the range of 10\,Hz to 1\,MHz.}
\label{fig_phasenoise}
\end{figure}

\begin{figure}[ht]
\centering
\includegraphics[width=\textwidth]{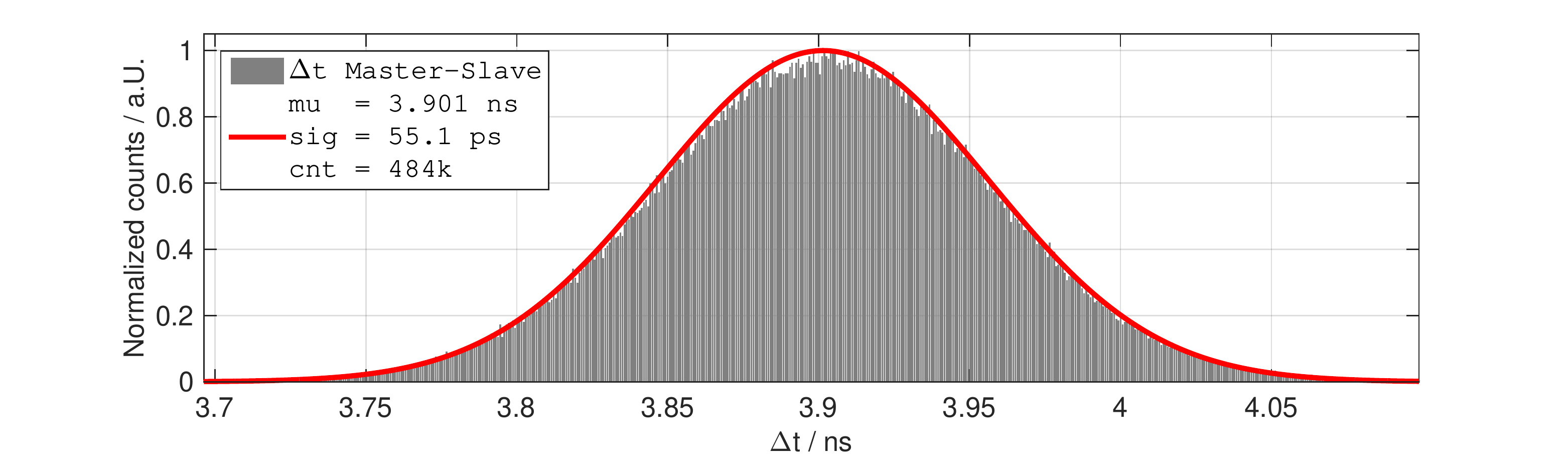}
\caption{Clock to clock jitter between the master and the slave PHY. Both are the DP83865. The standard deviation of the distribution is 55.097\,ps.}
\label{fig_deltat_clock_ti}
\end{figure}

\begin{figure}[ht]
\centering
\includegraphics[width=0.75\textwidth]{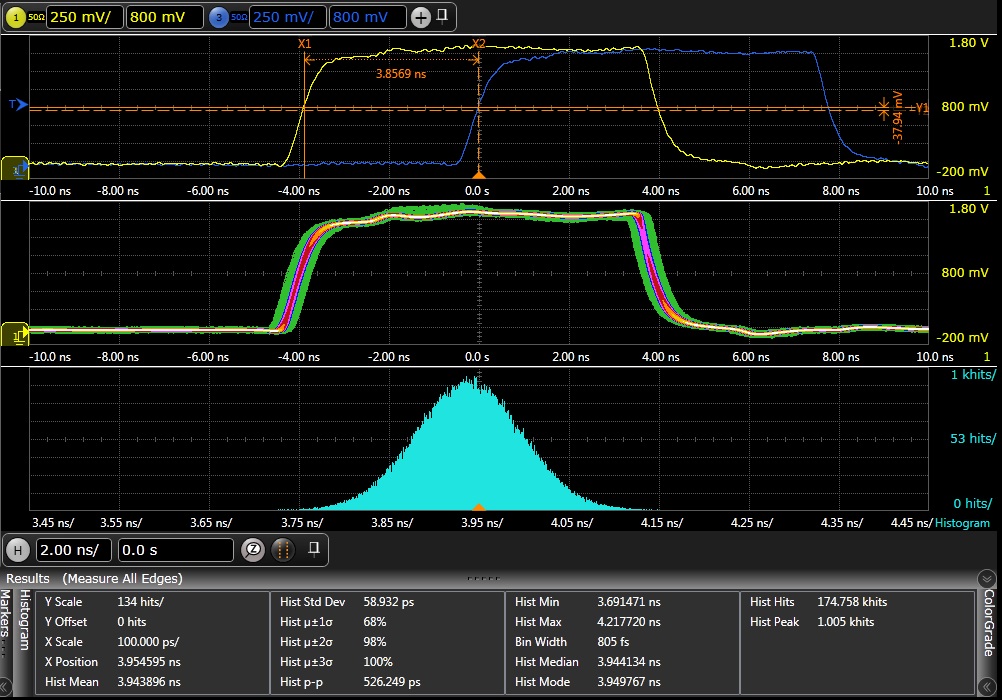}
\caption{Oscilloscope measurement for the time difference between the master's and the slave's timestamps. The FPGA outputs a PPS signal (for this measurement every 268.4\,ms) which shows the precision of the absolute timestamp synchronization with PTP. The standard deviation of the time difference is 58.932\,ps with a constant offset of about 3.94\,ns}
\label{fig_pps}
\end{figure}

The performance of the clock synchronization is limited by the accuracy of the clock recovery system in the signal chain of the 1000BASE-T slave. Whereas the master's clock can achieve the desired precision by choosing an appropriate clock source, the precision of the slave's clock depends on its components in the signal chain for the clock distribution and recovery (see fig.~\ref{fig_clocking_scheme}). For our evaluation hardware, the accuracy of the signal chain is mainly determined by the PHY which is responsible for the clock recovery out of the datastream. The second component which influences the absolute precision is the FPGA, where the recovered clock is used for timestamp generation. Usually a PLL inside the FPGA is used to build the clock tree for all clock domains. So we want to evaluate, whether the built-in PLL limits the overall system. At first we measured the phase noise of a clock source with very low jitter which will be used as the input signal for the PLL. The phase noise is correlated with the random jitter and therefore it determines the precision of the timing system. The measurements of a low jitter clock and the performance of the FPGA's (Spartan~6 LX45T) built-in PLL with that input is shown in fig.~\ref{fig_phasenoise}. These measurements were taken with a HA7062B phase noise analyzer from Holtzworth Instrumentation and the signal generator SMA100A from Rohde \& Schwarz as low jitter clock source.
Although there are various configurations possible for the PLL, for this measurement we set up the multiplier and the divider value to 8. The integrated phase noise of the Xilinx PLL was found out to be 6.47\,ps in the range of 10\,Hz to 1\,MHz. Without any additional hardware, this constitutes a design limit for the precision of synchronous timestamp generation with the FPGA. To find out the random jitter of the clock synchronization over a 1000BASE-T link, we set up a point-to-point link between a master and a slave PHY and measured the clock to clock jitter in the time domain. The master's clock triggers the measurement of time difference between the two rising edges of both clocks. The results of the measurement for the PHY DP83865 (master and slave) are shown in fig.~\ref{fig_deltat_clock_ti}.
The clock signal is measured at an output pin of the FPGA with a frequency of 125\,MHz (see fig.~\ref{fig_clocking_scheme}). It is buffered with an output register (ODDR2 primitive from Xilinx). During the measurement in fig.~\ref{fig_deltat_clock_ti} the FPGA sends UDP packets with maximum throughput and PTP packets at an interval of 1\,s. We also bypassed the PLL and distributed the recovered slave clock with an ordinary built-in clock buffer to the timing logic. With regard to that, the jitter was increased from 55.097\,ps to 64.273\,ps. Finally, we repeated the measurement of fig.~\ref{fig_deltat_clock_ti} with the PHY 88E1111 for the master and the slave. With this setup we achieved a clock to clock jitter of 70.33\,ps. In both setups the master's clock source was an crystal oscillator with approximately 6\,ps random jitter (measured with the phase noise analyzer in the range from 10\,Hz to 1\,MHz). As a result, we can state that the precision is influenced by all components in the signal chain. It depends mainly on the clock recovery system of the PHY and the ability of the FPGA's PLL to reduce random jitter.
In addition to the measurement of the clock to clock jitter, we have taken measurements to estimate the synchronization of the master's and slave's timestamps. Both devices run on the synchronized clock signal with the same frequency. An absolute synchronization of the timestamps is performed with PTP every second. Each synchronized device generates one pulse per second (PPS) at an output of the FPGA. A measurement of the time difference between the PPS signal of the master and the slave is shown in fig.~\ref{fig_pps}.
The measurement was running over 13 hours in the lab and shows that the timestamps are synchronized with a random jitter of 58.932\,ps. The digital logic for timestamp generation in the FPGA was sourced by a clock signal with a frequency of 125\,MHz from the built-in PLL. For the measurements shown in fig.~\ref{fig_pps} we used the DP83865. The same measurements were repeated with the 88E1111 PHY and resulted in a random jitter of 71.536\,ps for the timestamp synchronization in a short-term measurement (approx. 2\,hours). All measurements showing a constant offset up to 8\,ns which cannot be reduced with the PTP. Further investigations have to be done with excessive temperature stress for the PHYs and the FPGAs.

\subsection{Resource utilization}
Our Ethernet IP core can be configured with an arbitrary FIFO size for the application above the UDP layer. Our basic configuration consists of two channels for the application interface to the UDP layer. One channel is interfaced by the microcontroller and one is interfaced by a high-throughput application. On each interface there is one FIFO for the payload with a depth of at least two UDP packets. The payload size of one packet is set by generics and is by default 1472\,Byte. For the support of jumbo frames this size can be easily adjusted to 8972\,Byte. Larger FIFO depths and payload sizes are possible as well. The FIFOs are implemented on Dual-Port Block Memory integrated in the Xilinx FPGA. They can also be placed on distributed slice registers. The ICMP, PTP and ARP layer can store one packet to send. All receiving datapaths are configured to store one packet as well. Because there are various configurations possible, we renounce a comparison to other implementations. The resource utilization reported by the Xilinx tools for the Ethernet stack with support for ARP, PTP, ICMP and UDP is presented in tab.~\ref{tab_util_spartan} and tab.~\ref{tab_util_kintex}, using the Xilinx ISE 14.7 tools for the implementation.
\begin{table}[ht]
\caption{Slice Logic utilization for the Gigabit Ethernet stack with a Spartan~6 (LX45T)}
\label{tab_util_spartan}
\centering
\begin{tabular}{|l|l|l|l|l|l|}
\hline
\bfseries Module & \bfseries Slices & \bfseries Slice Reg & \bfseries LUTs & \bfseries LUTRAM & \bfseries BRAM\\
\hline				
MAC      & 140	& 459	  & 345	  & 16	& 0\\
\hline					
Ethernet & 246	& 479	  & 648	  & 0	  & 0\\
\hline	
ARP      & 163	& 498	  & 492	  & 0	  & 0\\				
\hline	
IP       & 274	& 546	  & 669	  & 0	  & 0\\
\hline	
ICMP     & 60 	& 169	  & 108	  & 24	& 1\\
\hline	
UDP      & 237	& 551	  & 573	  & 1	  & 9\\
\hline	
PTP      & 672	& 1890	& 2071	& 0	  & 0\\
\hline
\hline
{Sum}         & 1792 & 4592 & 4906 & 41 & 10\\
\hline
\end{tabular}
\end{table}
The 1000BASE-T implementation with support for PTP and a MTU of 1500\,Byte on a Spartan~6 FPGA with 6822 slices occupies 1792 slices corresponding to 26.27\,\% (16,42\,\% without PTP) total occupied slices.\\
The implementation on Kintex~7 is designed for a 1000BASE-KX link on a MicroTCA backplane. This implementation uses a Xilinx IP core with a GTX transceiver as PHY. This consumes additional logic but doesn't need an external PHY. This implementation aims only at maximum data throughput and is not designed to perform a synchronization over the MicroTCA backplane. Thus, the PTP layer is not included.
\begin{table}[ht]
\caption{Slice logic utilization for the Gigabit Ethernet stack with a Kintex~7 (325T)}
\label{tab_util_kintex}
\centering
\begin{tabular}{|l|l|l|l|l|l|}
\hline
\bfseries Module & \bfseries Slices & \bfseries Slice Reg & \bfseries LUTs & \bfseries LUTRAM & \bfseries BRAM\\
\hline
GMII\_to\_GTX & 446	& 997	& 826	& 71	& 0\\
\hline				
MAC           & 94	& 299	& 276	& 32	& 0\\				
\hline
Ethernet      & 137	& 378	& 419	& 0	  & 0\\
\hline
ARP           & 173	& 498	& 485	& 0	  & 0\\				
\hline
IP            & 250	& 546	& 684	& 0	  & 0\\
\hline
ICMP          & 61	& 169	& 114	& 24	& 1\\
\hline
UDP           & 198	& 580	& 580	& 1	  & 5\\
\hline
\hline
{Sum}         & 1359 & 3467 & 3384 & 128 & 6\\
\hline
\end{tabular}
\end{table}
The 1000BASE-KX implementation without support for PTP and a MTU of 1500\,Byte on a Kintex~7 FPGA with 50959 Slices occupies 1359 Slices corresponding to 2.67\,\%.\\
All reports for slice logic utilization also include several logic for internal tests and debug options.

\section{Summary}
With the need of a high-throughput UDP application, we have presented an entire stack architecture for a Gigabit Ethernet interface on a FPGA. The stack was built for the protocols UDP, ICMP, IP, ARP and PTP and can be easily extended or cut down in functionality. For a straight forward implementation we showed two basic models for the dataflow in a stacked architecture. Our embedded Gigabit Ethernet protocol stack is designed with the "Data-Pull" model to eliminate redundant buffers. A clear modular architecture for each layer with a control and arbiter logic at the interconnections keeps this implementation versatile. The underlying MAC and physical layer are also replaceable. All modules are written in VHDL and tested on Xilinx Spartan~6 and Kintex~7.
\\
We demonstrated the data throughput with an UDP application on a {1000BASE-T} and {1000BASE-KX} link. In both cases we achieved the maximum data throughput of 114.1\,MiB/s with a MTU of 1500\,Byte and 118.3\,MiB/s with jumbo frames of 9000\,Byte. The overall performance for other use cases is also excellent. Finally, we investigated the performance of a clock synchronization over a 1000BASE-T link. In dependence of the PHY, we achieved a precision of 55.1\,ps for the clock to clock jitter between the master and the slave. An absolute synchronization of timestamps was done with PTP. The long-term test showed a standard deviation of 58.9\,ps for the synchronized timestamps. Due to the generic data interface, this UDP/IP stack can be easily adapted to detector applications where high data throughput is required. For precise timing applications the relative timing is in the sub nanosecond range whereas the absolute accuracy remains in the limits of PTP.

\end{document}